\documentclass[pra,twocolumn,showpacs,amsmath,amssymb] {revtex4}

\usepackage{color}
\usepackage{graphicx}

\begin{document}

\title{Dynamics of spin-1 bosons in an optical lattice: spin mixing, quantum phase revival
spectroscopy and effective three-body interactions}

\author{K. W. Mahmud and E. Tiesinga}

\affiliation{Joint Quantum Institute, National Institute of Standards
and Technology and University of Maryland, 100 Bureau Drive, Stop 8423,
Gaithersburg, Maryland 20899, USA}

\begin{abstract}
We study the dynamics of spin-1 atoms in a periodic optical-lattice
potential and an external magnetic field in a quantum quench scenario
where we start from a superfluid ground state in a shallow lattice
potential and suddenly raise the lattice depth. The time evolution of the
non-equilibrium state, thus created, shows collective collapse-and-revival
oscillations of matter-wave coherence as well as oscillations in
the spin populations.  We show that the complex pattern of these two
types of oscillations reveals details about the superfluid and magnetic
properties of the initial many-body ground state.  Furthermore, we show
that the strengths of the spin-dependent and spin-independent atom-atom
interactions can be deduced from the observations.  The Hamiltonian that
describes the physics of the final deep lattice not only contains two-body
interactions but also effective multi-body interactions, which arise
due to virtual excitations to higher bands.  We derive these effective
spin-dependent three-body interaction parameters for spin-1 atoms and
describe how spin-mixing is affected.  Spinor atoms are unique in the
sense that multi-body interactions are directly evident in the {\it
in-situ} number densities in addition to the momentum distributions.
We treat both antiferromagnetic (e.g. $^{23}$Na atoms) and ferromagnetic
(e.g. $^{87}$Rb and $^{41}$K) condensates.

\end{abstract}


\maketitle


\section{Introduction}

Quantum degenerate ultracold atoms with spin-degree of freedom
exhibit both magnetic order and superfluidity, offering a rich
system in which to explore quantum coherence, long-range order,
magnetism and symmetry breaking.  Many aspects of spinor atoms in
a trap have been investigated with spin $F=1$ atoms, such as
$^{23}$Na and $^{87}$Rb
\cite{ketterle98,lett09,vengalattore08,dalibard12,raman13}. Spin-2
\cite{schmalljohann04,chang04,kuwamoto04} and spin-3
\cite{Santos2006,pasquiou11} spinor gases have been studied to a
lesser extent.  Spinor condensates are described by a vector order
parameter~\cite{ho98,kurn12,ueda12}.  The distinctive feature is
its spin-dependent interaction which organizes spins giving rise
to ferromagnetic and antiferromagnetic (polar) order. It can also
coherently convert a spin $m=1$ and a $m=-1$ atom to two $m=0$
atoms and vice versa~\cite{law98,chapman05,chapman05b}, while
conserving magnetization and energy.

In parallel to ultracold spinor physics, optical lattices have
become a powerful tool to create strongly correlated many-body
states of bosons and
fermions~\cite{jaksch98,greiner02a,bloch08,lewenstein07,sengstock12}.
Lattice systems offer flexibility as the lattice parameters and
particle interactions can be controlled easily, thereby
facilitating progress towards the creation of quantum
emulators~\cite{spielman10,mahmud11}.  Since the seminal
observation of the superfluid to Mott insulator transition with
spinless bosons~\cite{greiner02a}, steady progress is being made
towards the understanding of spinor atoms in an optical
lattice~\cite{widera06,trotzky08,sengstock10}. Issues of
temperature and entropy~\cite{mahmud10} are among the challenges
that need to be overcome to create a many-body correlated state of
spin-1 atoms. Theoretical studies of lattice-trapped spinor
condensates have mainly explored the phase diagram and the nature
of the superfluid-Mott insulator transition
\cite{demler02,snoek04,imambekov03,batrouni09,pai08,santos11}.

Due to the tunability of cold atom and optical lattice parameters,
it is also possible to study non-equilibrium dynamics. Dynamics
of many-body quantum systems is still an emerging field, and only a
number of issues have so far been investigated~\cite{polkovnikov11}. An
early experiment~\cite{greiner02b} studied the dynamics of spinless
bosons in a suddenly-raised optical lattice, observing the collapse
and revival of the matter wave field in the momentum distribution. In
a more-recent experiment \cite{will10}, tens of oscillations
in the momentum distribution or visibility were observed, and the
predicted~\cite{tiesinga09} signature of effective higher-body
interactions confirmed. As for spin-1 atoms, dynamical studies have
mainly focused on large atom continuum or trapped systems in the mean
field regime exploring spin-mixing dynamics~\cite{law98,chapman05},
quantum quench dynamics~\cite{kurn06,kurn07}, and various
instabilities~\cite{chapman11}. 

The goal of this paper is to study the dynamics of spin-1 bosonic
atoms in a three-dimensional (3D) optical lattice and probe its
many-body state and system properties. Starting with a
ferromagnetic ($^{87}$Rb) or antiferromagnetic ($^{23}$Na)
superfluid ground state in a shallow lattice, suddenly raising the
lattice depth creates a non-equilibrium state, which can be
followed in various scenarios -- with and without a magnetic field
and with and without effective three-body interactions. The
evolution shows collapse and revival of matter-wave coherence
measured by visibility oscillations, in a more complex pattern
than for spinless bosons~\cite{greiner02b,porto07}. It also shows
oscillations in spin populations due to the combined effect of the
spin-mixing collisions of the $m=0$ and $m=\pm 1$ components and
differential level shifts proportional to the square of the
magnetic field strength, the quadratic Zeeman shift.  Linear
Zeeman shifts do not affect the behavior of spinor condensates.
Both spin-mixing and visibility oscillations reveal details about
the system such as the composition of the initial many-body state,
and thereby its superfluid and magnetic properties.

By analyzing the frequency spectrum of the visibility, we show
that the ratio $U_2/U_0$ of spin-dependent and spin-independent
atom-atom interactions can be deduced. Combined with spectra of
spin-mixing dynamics at various magnetic field strengths, this
gives us a method to measure the interaction couplings for spin-1
atoms.  Finally, we find that the presence of quadratic Zeeman
shift enhances spin mixing oscillations for ferromagnets and shows
collapse and partial revivals in the transverse magnetization.

The Hamiltonian that quantitatively describes the physics of the
final deep lattice comprises of two-body as well as effective
multi-body interactions, which arise due to virtual excitations to
higher bands. We derive the induced three-body interaction
parameters for spin-1 atoms in a deep harmonic well, approximating
the minimum of a single lattice site as such, and find the
existence of {\it spin-dependent} three-body interactions.  We
show how to detect the signature of the three-body interactions
and argue that they are directly exhibited in the {\it in situ}
density as opposed to the time of flight visibility measurements
as is the case for spinless bosons~\cite{will10}.

The article is organized as follows.  In Sec.~\ref{sec:model} we
setup the spin-1 Bose-Hubbard model, sketch the mean-field theory
to determine the initial ground state, describe the exact
Hamiltonian after the quench, and discuss observables and
computational aspects. We present our main results in
Sec.~\ref{sec:polar}, \ref{sec:ferro}, and \ref{sec:three}.
Section~\ref{sec:polar} explores the non-equilibrium dynamics of
antiferromagnetic spin-1 atoms, with and without a magnetic field.
Section~\ref{sec:ferro} describes the dynamics for a ferromagnetic
spinor. Section~\ref{sec:three} shows the effects of effective
three-body interactions in the dynamics. We summarize our results
in Sec.~\ref{sec:conclusion}. A derivation of the effective
three-body interaction is given in the appendix.

\section{Model and computational aspects} \label{sec:model}

\subsection{Shallow Lattice Hamiltonian} \label{sec:shallow}

Ultracold spin-1 bosons in the lowest band of a 3D cubic
optical lattice and an external magnetic field $B$ along the $z$ axis
are modeled by the free energy
\begin{eqnarray}
 H &=& -J \sum_{\langle i,j \rangle,m} \left( a^\dagger_{im}
a^{}_{jm} + a^\dagger_{jm} a^{}_{im} \right)
+ \frac{U_0}{2} \sum_i \hat{n}_{i} \left( \hat{n}_{i} -1 \right) \nonumber \\
&&
 + \frac{U_2}{2} \sum_i \left( \vec{\cal F}_{i} \cdot \vec{\cal F}_{i}
                                - 2\hat{n}_{i}\right)
 + \delta \sum_{im}  m^2 a^\dagger_{im} a_{im}
 \nonumber\\
&&\quad\quad - \mu  \hat{N} - \mu_M \hat M \,.
\label{hamil}
\end{eqnarray}
Here $a^\dagger_{im}$ is the creation operator of a boson in
magnetic sublevel $m = -1, 0$, or $1$ in the energetically-lowest
Wannier function or orbital of lattice site $i$. The first term in
Eq.~(\ref{hamil}) represents the hopping of atoms between nearest
neighbor sites and is proportional to the hopping energy $J$. The
second term describes the on-site atom-atom repulsion with
strength $U_0>0$, $\hat{n}_i = \sum_m \hat{n}_{im}$, and
$\hat{n}_{im}= a^\dagger_{im} a_{im}$.  The third term is a
spin-dependent atom-atom interaction with a strength $U_2$ that
can be either positive or negative. The three operators $\vec{\cal
F}_{i}=(\hat{\cal F}_{xi}, \hat{\cal F}_{yi},\hat{\cal F}_{zi})$
on site $i$ satisfy angular momentum commutation rules and are
defined by $\hat{\cal F}_{q i}=\sum_{m,m'}a^\dagger_{im} (F_q)_{m
m'} a_{im'}$ for $q=x$, $y$, or $z$ and $(F_q)_{m m'}$ are matrix
elements $m$, $m'$ of component $\alpha$ of the spin-1 angular
momentum $\vec F$. The fourth term corresponds to the quadratic
Zeeman energy of the magnetic sublevels with strength $\delta$.
Finally, the terms containing the Lagrange multipliers $\mu$ and
$\mu_M$ control the total atom number, $\hat N=\sum_i \hat{n}_i$,
and total magnetization, $\hat M= \sum_i {\hat m}_i$,
respectively.  Here ${\hat m}_i\equiv \hat{\cal F}_{zi}= \sum_m m
\,{\hat n}_{im}$ is the on-site magnetization.  Both $\hat N$ and
$\hat M$ commute with $H$.  (The large linear Zeeman Hamiltonian
of the atoms is ``absorbed'' in the term $- \mu_M \hat M$ and seen
not to affect the physics of the spinor condensate.)

The interaction strengths are given by $U_0=4\pi \hbar^2 {\bar n}
(a_0+2a_2)/(3M_a)$ and $U_2=4\pi \hbar^2 {\bar n}
(a_2-a_0)/(3M_a)$ \cite{ho98}, where $a_F$ with $F=0$ or 2 are
scattering lengths for the two allowed values of the total angular
momentum of $s$-wave collisions of two spin-1 particles at zero
collision energy and zero magnetic field. $S$-wave scattering with
total angular momentum $F=1$ is prohibited due to bosonic wave
function symmetry. The mean density of the local orbital ${\bar
n}$ is determined by the laser parameters and polarizability of
the atom. Finally, $M_a$ is the mass of the atom,
$\hbar=h/(2\pi)$, and $h$ is Planck's constant.

The ratio of the two interaction strengths is independent of
lattice parameters as the ${\bar n}$ dependence cancels. For
$^{23}$Na $U_2/U_0=+0.036(3)$~\cite{Knoop2011} and the system is
antiferromagnetic. For $^{87}$Rb
$U_2/U_0=-0.0046(7)$~\cite{vanKempen2002} and the system is
ferromagnetic. The quoted uncertainty in $U_2/U_0$ for $^{23}$Na
and $^{87}$Rb are one-standard deviation and obtained from the
corresponding references.  For $^{41}$K we find $U_2/U_0= -0.012$
\cite{Falke2008,Lysebo2010}.  For our investigation we use
$U_2/U_0=0.04$ for sodium and $-0.005$ for rubidium.  The
quadratic Zeeman strength $\delta=\delta_0 B^2$ with
$\delta_0/h=27.68 $ Hz/(mT)$^2$ for $^{23}$Na and
$\delta_0/h=7.189 $ Hz/(mT)$^2$ for $^{87}$Rb.

The phase diagram for the spin-1 Bose-Hubbard model has been
calculated with numerical methods such as Quantum Monte
Carlo~\cite{batrouni09} and density-matrix renormalization
group~\cite{rizzi05} for a one-dimensional lattice. Mean-field
approaches for spin-1 bosons, which give predictions for any
dimension, have also been
performed~\cite{pai08,krutitsky04,kimura05} and are extensions of
those for scalar bosons~\cite{sheshadri93,stoof01}. As mean-field
models are most predictive in three dimensions and we focus on
such a lattice, we use the decoupling mean-field
theory~\cite{pai08} to find the initial many-body ground state.

In a mean-field approximation, the hopping term in Eq.~(\ref{hamil}) can be
decoupled as
\begin{equation}
a^\dagger_{im} a_{jm} \simeq \langle a^\dagger_{im}\rangle a_{jm}+
a^\dagger_{im} \langle a_{jm} \rangle-\langle
a^\dagger_{im}\rangle \langle a_{jm} \rangle \, , \label{mfa}
\end{equation}
when fluctuations around the equilibrium value are negligible. We can
define $\psi_{m}= \langle a_{jm} \rangle$, for $m=1, 0, -1$, as the
site-independent superfluid order parameter. Using Eq.~(\ref{mfa})
we can rewrite Eq.~(\ref{hamil}) as a sum of independent single site
Hamiltonians, $H=\sum_{i} H^{\rm mf}_{i}$ where
\begin{eqnarray}
\lefteqn{H^{\rm mf}_{i} = \dfrac{U_0}{2} \hat{n}_{i} \left( \hat{n}_{i}
-1 \right) + \dfrac{U_2}{2} \left( \vec{\cal F}_i\cdot \vec{\cal F}_{i}
-2\hat{n}_{i}\right) }  \nonumber\\
 && + \delta \sum_{m}  m^2 a^\dagger_{im}  a_{im}
- \mu \hat{n}_i -\mu_M \hat{m}_i
     \label{mfham} \\
&& - z J\sum_{m} (\psi_{m} a^\dagger_{im} + \psi_{m}^*
a_{im} ) + z J\sum_{m} |\psi_{m}|^2\,, \nonumber
\end{eqnarray}
and $z$ is the number of nearest neighbors, e.g. $z=6$ in 3D. For
a given $\mu$ and $\mu_M$ the superfluid order parameters and
ground state wavefunction are obtained by finding those values of
$\psi_{m}$ for which the energetically-lowest eigenstate of
$H^{\rm mf}_{i}$ is smallest.

The character of the ground state depends
on whether the spin-dependent term $U_2$ is positive or negative
\cite{ho98}.  For the antiferromagnetic $U_2>0$ superfluid ground
states, the order parameters can be written as $\psi_m=\sqrt{\rho_s}
e^{i\theta}D^1_{m0}(\alpha,\beta,\gamma)$, while for  the
ferromagnetic $U_2<0$ superfluid we have $\psi_m=\sqrt{\rho_s}
e^{i\theta} D^1_{m1}(\alpha,\beta,\gamma)$.  Here the functions
$D^J_{MM'}(\alpha,\beta,\gamma)$ are Wigner rotation matrices
\cite{brink93} with Euler angles $\alpha$, $\beta$, and $\gamma$
determined by minimizing $H^{\rm mf}_i$.  The real valued $\rho_s$
and angle $\theta$ are the spin-independent superfluid density and a
global phase, respectively. We have $\rho_s\le
\langle {\hat n} \rangle$.

Within mean-field theory the many-body superfluid wavefunction is given
by the product wavefunction $\prod_i |{\rm GS}\rangle_i$ over sites $i$,
where $|{\rm GS} \rangle=\sum_{\vec n} c_{\vec n}|{\vec n}\rangle$
and kets $|{\vec n}\rangle=|n_{-1},n_{0},n_{1} \rangle$ are elements of the {\it
occupation-number} basis of Fock states of the three $m$ projections.
The single-site wavefunction is a superposition of Fock states with
amplitudes $c_{\vec n}$. In fact, it is also a superposition of Fock
states with different magnetization.

We only present results for ground states with zero magnetization
$\langle m_i \rangle=0$ at every site, ensured by setting
$\mu_M=0$. Results for other magnetizations show similar physics.
For $U_2>0$ the bosons condense into a state with $\langle
\vec{\cal F}_i \rangle=0$. This is called a polar
(antiferromagnetic) superfluid. There are two kinds of polar
order~\cite{pai08}: the ground state is a transverse polar state
with $(\psi_{-1},\psi_0,\psi_{1})=\sqrt{\rho_s}(1,0,1)/\sqrt{2}$
when $\delta<0$, and a longitudinal polar state with
$(\psi_{-1},\psi_0,\psi_{1})=\sqrt{\rho_s}(0,1,0)$ when
$\delta>0$. For ferromagnetic atoms with $U_2<0$ the magnetic
order maximizes the total angular momentum with $\langle \vec{\cal
F}_i \rangle^2=1$~\cite{ho98}, leading to a partially-magnetized
superfluid with order parameters
$(\psi_{-1},\psi_0,\psi_{1})=\sqrt{\rho_s}(1,\sqrt{2},1)/2$ for
$0<\delta<2U_2$, and a longitudinal superfluid with order
parameters $\sqrt{\rho_s}(0,1,0)$ for $\delta>2U_2$.

Our numerical simulations are performed in the occupation number
basis.  Only basis functions with $n_{-1}+n_{0}+n_{1} \leq n_{max}$ are
included. We use $n_{max}=6$ leading to 84 basis functions in a site and
negligible truncation errors when the mean atom number per site is less
than three. All current optical-lattice experiments
use mean atom numbers of this order of magnitude.

\subsection{Deep Lattice Hamiltonian} \label{sec:deep}

After preparing the initial superfluid the depth of the optical
lattice is suddenly increased so that tunneling is turned off.
This lattice ramp-up is assumed to be slow enough to prevent
excitations to a higher band yet fast enough compared to
interactions.  We can then treat subsequent time evolution due to
the single-site Hamiltonian
\begin{equation}
H^{\rm final} =
  \dfrac{U_0}{2} \hat{n} \left( \hat{n} -1 \right)
+ \dfrac{U_2}{2} \left( \vec{\cal F}\cdot\vec{\cal F} -2\hat{n}\right)
  + \delta \sum_{m}  m^2a^\dagger_{m}  a_{m}
\label{final}
\end{equation}
exactly. As each site evolves under the same Hamiltonian, we have
suppressed the site index. For induced three-body interactions,
additional terms appear in $H^{\rm final}$ as discussed in
Sec.~\ref{sec:three}. Recent observation of multi-body effects for
lattice-trapped spinless bosons~\cite{will10}, where a similar
quench was used, a mean-field
treatment~\cite{tiesinga09,tiesinga11} of the initial state
followed by exact on-site evolution was found to agree well with
the experiment. If lattice sites are not completely decoupled ($J
\neq 0$) during the evolution, a correlated multi-site treatment
is necessary~\cite{daley11}. We do not study that scenario in this
paper.

Following Ref.~\cite{law98} we realize that, in addition to the
occupation number basis, eigenfunctions of the operators $\hat n$,
$\vec{\cal F}^2=\vec{\cal F}\cdot\vec{\cal F}$, and ${\cal F}_z$
also form a complete basis for the on-site Hilbert space of
$H^{\rm final}$. In fact, these {\it angular-momentum} basis
states  $|n,F,M \rangle$ diagonalize $H^{\rm final}$ when
$\delta=0$ with energy spectrum
\begin{equation}
E^{\rm final}(\delta=0) = \dfrac{U_0}{2} n ( n-1 )+ \dfrac{U_2}{2} \left( F(F+1) -2n\right)\,,
\label{energy}
\end{equation}
where $n$ is the local atom number and the quantum number $F$ is restricted to
$F\le n$ and even/odd $F$ for even/odd $n$.  The integer $M$ is the
magnetization quantum number with $|M|\le F$.
For $\delta\ne 0$ the quadratic Zeeman interaction couples the
angular-momentum basis states.  Finding its matrix elements is involved,
leading us to perform all simulations in the occupation number basis.

\subsection{Observables}

To analyze the non-equilibrium dynamics of our system, we follow
several observables. The first is the atom number per lattice site
in each spin component, $\langle \hat{n}_{im}\rangle$, which can
be detected either in situ~\cite{greiner09} or, after release of
the atoms from the lattice, by the Stern-Gerlach separation method
where the spin states are first spatially separated and then
detected~\cite{chapman05b}. The second observable is the
visibility, which is a measure of coherence of the wavefunction
and equals the number of atoms with zero momentum in the
spin-dependent momentum distribution.  This is the standard
quantity measured after releasing atoms from the lattice and a
time of flight expansion~\cite{will10,will11}.  Within our
simulation it is given by $|\langle \hat{a}_{im} \rangle|^2$ for
any site $i$. Finally, we study the square of the in-situ
transverse magnetization $|\langle \hat{\cal F}_{ix} \rangle|^2$.
Transverse magnetization can be measured by Faraday rotation
spectroscopy which allows for continuous observation of spin
population in a BEC~\cite{lett09,lett07}.

\section{Dynamics of an antiferromagnetic spinor} \label{sec:polar}

\subsection{Evolution without magnetic field}

\begin{figure}
\vspace{-0.1cm}
\begin{center}
  \includegraphics[width=0.4\textwidth,angle=0]{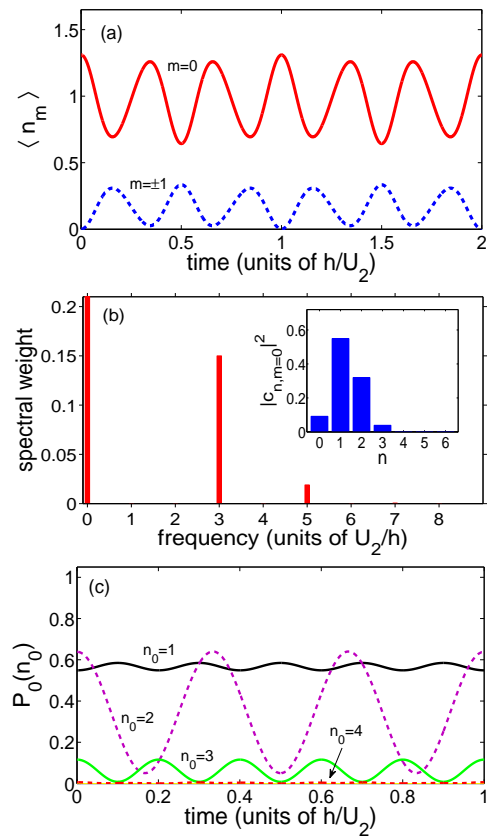}
\end{center}
\vspace{-1.5cm} \caption{\label{fig:mixingAF} (color online)
Spin-mixing dynamics after a sudden increase of lattice depth
starting with the longitudinal antiferromagnetic ground state of
spin-1 $^{23}$Na in a zero magnetic field.  (a) The on-site atom
number of the spin components, $\langle {\hat n}_{m=0,\pm 1}
\rangle$, as a function of hold time showing spin-mixing.  Time is
in units of $h/U_2$ and system parameters are described in the
text. (b) Fourier spectrum of the time trace in (a) showing the
frequencies involved in the spin-mixing dynamics. The inset shows
the initial ground state Fock state number probabilities
$|c_{{\vec n}=(0,n_0,0)}|^2$ as a function of atom number $n_0$ in
spin component $m=0$. There are no atoms with $m=\pm 1$. (c)
Contribution $P_0$ to $\langle {\hat n}_{m=0} \rangle$ of having
$n_0$ atoms in the $m=0$ state, where $P_0(n_0)=\sum_{n_{m=\pm 1}}
n_0 |c_{\vec n}|^2$. The curves are for $n_0=1,2,3$, and $4$. It
shows that the $3h/U_2$ periodicity is due to $n_0=2$ and $5h/U_2$
periodicity due to $n_0=1$ and 3, respectively.}
\end{figure}

In this section we analyze the dynamics of a longitudinal polar
superfluid ground state after the optical lattice strength is
rapidly raised. The condensate evolves under $H^{\rm final}$ with
$\delta=0$ for hold time $t$, after which one or more of the
observables is measured. Figure~\ref{fig:mixingAF}(a) shows
typical evolution of the in-situ population of spin components
$m=\pm 1$, $0$ as a function of hold time. Here we use typical
numbers for $^{23}$Na atoms -- in the initial lattice with
$U_0/(zJ)=2$, $U_2=+0.04 U_0$, and a mean atom-number per site of
$\langle {\hat n} \rangle=1.31$, and in the final lattice $U_0=0.1
\hbar \omega_f, U_2=+0.04 U_0$, where $\omega_f$ is the harmonic
frequency near the minima of the lattice potential. (An
inifinitesimal $\delta>0$ is applied to ensure formation of the
longitudinal polar state.)

Initially, all atoms are in the $m=0$ state and as time evolves
atoms begin to appear in states $m=\pm 1$ because of spin changing
collisions, and a pattern of periodic modulation emerges with a
period of $h/U_2$, while conserving zero magnetization. This
spin-mixing time trace also contains information about the
composition of the initial many-body state. To explore this, we
further analyze the dynamics in Fig.~\ref{fig:mixingAF}(b) and
(c). Figure \ref{fig:mixingAF}(b) shows the Fourier analysis of
the time trace in panel (a), and the inset shows on-site Fock
state probabilities $|c_{\vec n}|^2$ for the initial state. The
shape of the initial state number distribution is characteristic
of a nearly coherent or a slightly-squeezed state. In the
frequency spectrum, peaks are observed at frequencies that are
integer multiples of $U_2/h$. In fact, they occur at $3U_2/h$,
$5U_2/h$, and a small contribution at $7U_2/h$. These features can
be understood from an analysis of the eigenenergies in
Eq.~(\ref{energy}) of $H^{\rm final}$ at $\delta=0$. Similar to
number Fock state composition of the initial state, it is also a
superposition of angular momentum states $|n,F,M\rangle$. The
observables ${\hat n}_m$ commute with total atom number $\hat n$
and thus only measure the coherence between states with different
$F$ but the same $n$. For states with $n=2$ the two allowed $F$
are $0$ and $2$ with energy difference $3U_2$. This leads to a
peak at $3 U_2/h$ in Fig.~\ref{fig:mixingAF}(b). For $n=3$ states,
$F=1$ and $F=3$ exist leading to the frequency at $5 U_2/h$. The
small feature at $7 U_2/h$ indicates the presence and mixing of
$|n,F,M\rangle=|4,4,0 \rangle$ and $|4,2,0 \rangle$ states. The
above analysis of the eigen energies is confirmed in
Fig.~\ref{fig:mixingAF}(c). It depicts the time evolution of the
contribution $P_0$ to $\langle {\hat n}_{m=0} \rangle$ of having
$n_0$ atoms in the $m=0$ state, where $P_0(n_0)=\sum_{n_{m=\pm 1}}
n_0 |c_{\vec n}|^2$. Atom number $n_0=2$ has a period of
$(h/U_2)/3$ as we oscillate between states
$|n_{-1},n_0,n_1\rangle=|0,2,0\rangle$ and $|1,0,1\rangle$ with a
total of two atoms, while that for $n_0=1$ and $3$ has a period of
$(h/U_2)/5$. Here, we oscillate between the three atom states
$|n_{-1},n_0,n_1\rangle=|0,3,0\rangle$ and $|1,1,1\rangle$.

The spin-mixing dynamics can be compared and contrasted with other
spin-1 experiments. In Ref.~\cite{widera06}, a pair of $F=1$
$^{87}$Rb atoms were prepared in a single site of a deep optical
lattice in the Fock state $|0,2,0 \rangle$, and allowed to
spin-mix with $|1,0,1 \rangle$. Spin mixing oscillations between
two levels analogous to Rabi oscillations were observed with a
single frequency. On the other extreme, spin-mixing for a spinor
BEC with a large number of atoms has been discussed in theory and
observed in experiments~\cite{law98,chapman05}. They are in a
regime where a classical pendulum phase-space analysis is
appropriate~\cite{zhang05,Zhai2009}, and although there can be
spin-mixing oscillations for specific initial states, quantum
recurrences due to the discrete energy spectrum is absent. Our
analysis here explores the regime which is between these two --
the single Fock state regime and the regime of large atom number
condensate. As such, we are exploring a regime which can shed
light on the semi-classical transition to large condensate
dynamics, a topic for future investigation. In our analysis here,
we can analyze the multiple frequencies of the dynamics time trace
to probe the composition and atom number statistics of the initial
many-body state. 

\begin{figure}
\vspace{-0.1cm}
\begin{center}
  \includegraphics[width=0.45\textwidth,angle=0]{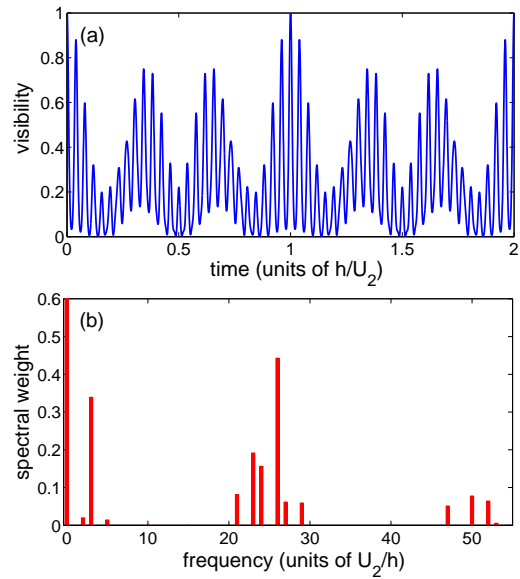}
\end{center}
\vspace{-1.7cm} \caption{\label{fig:visAF} (color online) (a)
Visibility $|\langle a_{m=0}(t) \rangle|^2$/$|\langle a_{m=0}(t=0)
\rangle|^2$ of the $m=0$ state as a function of hold time $t$
after a sudden increase of the lattice depth, for the same
parameters as in Fig.~\ref{fig:mixingAF}. The pattern of collapse
and revival of coherence is more complex than the spinless boson
case due to the competition between spin-dependent and
spin-independent interactions.  There are $U_0/U_2+1$ fast
oscillations in one full collapse and revival period of $h/U_2$,
yielding a method to determine the ratio $U_2/U_0$ from visibility
oscillations.  (b) Spectrum of the visibility oscillations showing
the range of contributing frequencies, the most dominant one being
at $U_0/U_2+1=26$.}

\end{figure}

Figure~\ref{fig:visAF} shows the dynamics of visibility of the
$m=0$ state -- the occupation of the zero momentum state for $m=0$
component, for the same initial state and parameters as in
Fig.~\ref{fig:mixingAF}. The visibility $|\langle a_{m=0}(t)
\rangle|^2$ measures the phase coherence in this spinor superfluid
system. We show the relative visibility $|\langle a_{m=0}(t)
\rangle|^2$/$|\langle a_{m=0}(t=0) \rangle|^2$ in
Fig.~\ref{fig:visAF}(a). We see that the atoms oscillate between
being completely coherent to completely incoherent ($|\langle
a_{m=0} \rangle|^2 \approx 0$). The pattern is more complex than
the spinless boson visibility~\cite{greiner02b}. There is a fast
oscillation with time scale $\sim h/U_0$, which is modified by a
slower envelope with a time scale $\sim h/U_2$. The exact nature
of the complex oscillations is revealed in the frequency spectrum
shown in Fig.~\ref{fig:visAF}(b). Similar to
Fig.~\ref{fig:mixingAF} features appear at small integer multiples
of $U_2/h$. Here they are located at $2U_2/h$, $3U_2/h$, and
$5U_2/h$. Peaks also occur at much larger frequencies with a
dominant frequency at $(U_0+ U_2)/h$, which is $26 U_2/h$ in this
example. Twenty six is also the number of fast oscillations in a
full period $h/U_2$ as can be seen in panel (a).  This indicates
that for an unknown ratio $U_2/U_0$, one full revival oscillations
of the visibility can help determine this ratio by counting the
number of fast oscillations or equivalently performing a frequency
analysis. Combined with the realization that this ratio is
independent of lattice parameters, this method of determining
spinor interactions is one of the key findings of this paper.

The visibility spectrum frequencies appear because the expectation
value of the annihilation operator is sensitive to the overlap
between Fock states of different atom number~\cite{will10}. For
example $|n_{-1},n_0,n_1\rangle=|0,2,0 \rangle$ connects to
$|0,1,0 \rangle$ giving rise to the dominant frequency $U_0+U_2$.
Other frequencies can similarly be explained by performing an
expansion of the initial state in the angular momentum basis, and
applying the evolution operator for the final Hamiltonian
Eq.~(\ref{final}) at $\delta=0$. For higher occupation numbers not
shown here we find that the visibility patterns become more
complex. In fact, by controlling the initial
squeezing~\cite{will10,tiesinga11} (i.e. by controlling the
initial tunneling energy $J$) as well as the total occupation, we
can change the complexity of the frequency spectrum and thereby
make it amenable for analysis.

\subsection{Effects of magnetic field}

The energies of three component spin-1 atoms are sensitive to
external magnetic fields. Such external fields have been exploited
in manipulating spinor atoms in an optical trap -- to access
ground state properties~\cite{ketterle98,lett09}, in detection
such as in Stern Gerlach separation method~\cite{chapman05b}, or
to influence the dynamics in quench experiments~\cite{kurn07}. For
spinor atoms in an optical lattice external magnetic fields cannot
be ignored and, in fact, lead to unique physics.  For example, the
quadratic Zeeman shift affects the phase diagram~\cite{santos11}.
The ratio of quadratic shift to the spin-dependent interaction
strength, $\delta/U_2$, controls the physics of this system. Two
different regimes emerge -- the Zeeman regime for $\delta>U_2$ and
the interaction regime for $\delta< U_2$~\cite{kurn12}.

\begin{figure}
\begin{center}
  \includegraphics[scale=0.7,trim=25 48 0 0,angle=0,clip]{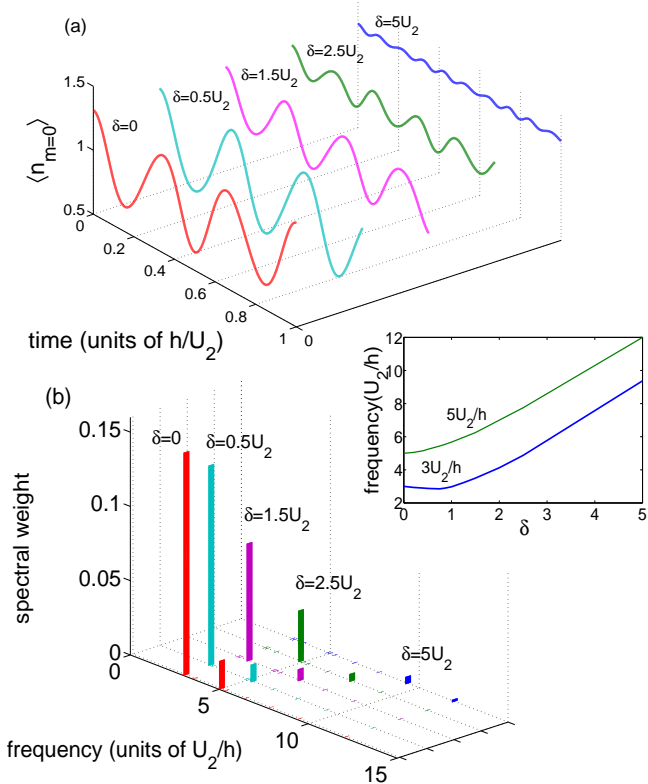}
\end{center}
\vspace{-0.2cm} \caption{\label{fig:afzeeman} (color online)
Spin-mixing oscillations in the presence of a magnetic field in
the form of a quadratic Zeeman shift. Panel (a) shows the
population in $m=0$ as a function of hold time for several values
of the quadratic Zeeman shift. Panel (b) shows the frequency
analysis of panel (a) with two dominant frequencies in the time
evolution.  At $\delta=0$ these frequencies are $3U_2/h$ and
$5U_2/h$. The inset shows that the two dominant frequencies as a
function of $\delta$ first dips before slowly increasing due to
the competing nature of $\delta$ and $U_2$.}
\end{figure}

Figure~\ref{fig:afzeeman} shows an analysis of spin-mixing
oscillations $\langle {\hat n}_{m=0}\rangle$ in the presence of a
quadratic Zeeman shift $\delta$, during the initial state
preparation and during the evolution. The other parameters are as
in Fig.~\ref{fig:mixingAF}. To highlight the effects of a magnetic
field, we show a comparison of the dynamics for $\delta/U_2=0,
0.5, 1.5, 2.5$, and $5$ in Fig.~\ref{fig:afzeeman}(a). We see that
the oscillations become faster while simultaneously the amplitudes
get smaller for increasing $\delta$ in the Zeeman regime
$\delta>U_2$. The spin-mixing dynamics vanishes for a large enough
$B$-field. In panel (b) we plot their frequency spectra comparing
the frequencies and the amplitudes.  The inset shows two dominant
frequencies as a function of the Zeeman strength. A closer look at
panel (b) reveals that for a $\delta$ in the interaction regime
$\delta<U_2$, the dominant frequency initially decreases before
starting to increase, due to a competition between the
spin-dependent interaction and the Zeeman term. Much of the
$B$-field effects can be understood from the eigenvalues of
$H^{\rm final}$ calculated in the Fock state basis~\cite{widera06}
using conservation of atom number and magnetization. For up to
three atoms per lattice site this involves diagonalizing at most a
$2\times 2$ matrix. The energy splitting of the eigenvalues of the
$2\times 2$ matrices is observed. In fact, from top to bottom the
two curves in the inset of panel (b) are due to contributions from
two-atom and three-atom Fock states. In a condensate with large
particle numbers the competition between interaction and Zeeman
energy gives rise to a sharp phase boundary at $\delta=U_2$ which
can be manifested through magnetic field induced
resonances~\cite{sengstock06}.

\section{Dynamics of a ferromagnetic spinor} \label{sec:ferro}

\subsection{Evolution without magnetic field}

\begin{figure}
\vspace{-0.1cm}
\begin{center}
  \includegraphics[width=0.5\textwidth,angle=0]{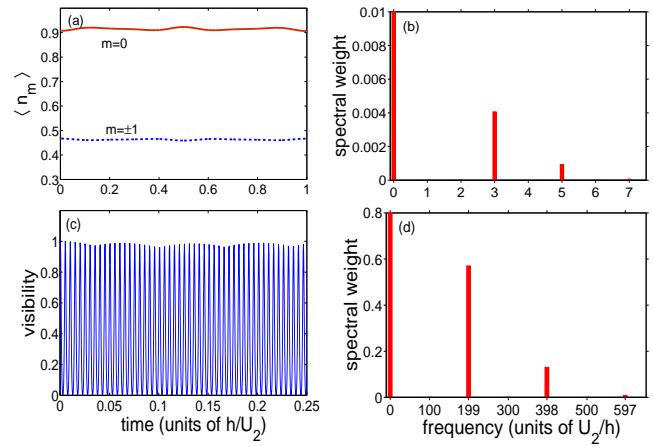}
\end{center}
\vspace{-1.0cm} \caption{\label{fig:ferro} (color online) (a)
Spin-mixing dynamics $\langle {\hat n}_m\rangle$ for a
ferromagnetic $U_2 <0$ rubidium condensate as a function of hold
time.  (b) Fourier spectrum showing the relevant frequencies of
3$U_2/h$, 5$U_2/h$ and 7$U_2/h$, and their relative amplitudes.
(c) Visibility $|\langle a_{m=0}(t)\rangle|^2$/$|\langle
a_{m=0}(t=0)\rangle|^2$ as a function of hold time. (d) Frequency
spectrum of the visibility showing features at $(U_0-U_2)/h$ and
$2(U_0-U_2)/h$, pointing a way to measure interaction ratios
$U_2/U_0$ from the visibility spectrum.}
\end{figure}

Figure~\ref{fig:ferro} shows the dynamics when our initial state
is a ferromagnetic superfluid state of $^{87}$Rb with $U_2<0$
created in a shallow lattice. For a ferromagnetic state, the
collective spin configuration is such that the spin-dependent
interaction energy is maximized. This means that the ground state
is a superposition of magnetization states in the angular momentum
basis $|n,F,M\rangle$ with variance $\Delta m \neq 0$ although it
still has magnetization $\langle {\hat m} \rangle=0$. The
parameters are $\langle {\hat n}\rangle =1.84$, $U_0/(zJ)=2$ and
$U_2=-0.005 U_0$ in the shallow lattice and $U_0=0.1\hbar\omega_f$
and $U_2=-0.005 U_0$ in the final deep lattice. The population
dynamics $\langle {\hat n}_m\rangle$ is shown in
Fig.~\ref{fig:ferro}(a). The oscillation amplitude is not as large
as in the polar case in the previous section. There are two
reasons for this: first, the ground state has comparable
populations in all three components and therefore, the population
difference between the spin components is smaller to begin with,
unlike for $^{23}$Na where initially only the $m=0$ state is
occupied. Second, the initial state is much closer to the ground
state in the deep lattice. We would like to point out that the
smallness of the ferromagnetic interaction is not responsible for
the oscillation amplitudes being small. A Fourier analysis in
panel (b) of the $m=0$ population shows that the frequencies
present are $3 U_2/h$, $5 U_2/h$ and $7 U_2/h$, as in the polar
state. Again, energy differences obtained from Eq.~(\ref{energy})
give us those frequencies. These frequencies and their spectral
weight determine the composition of different Fock components in
the initial many-body state, and therefore, can be used as an
experimental probe. We will show in the next subsection that
adding magnetic-fields can enhance the amplitude of spin-mixing
dynamics.

The visibility $|\langle a_{m=0}(t)\rangle|^2$/$|\langle
a_{m=0}(t=0)\rangle|^2$ is shown in Fig.~\ref{fig:ferro}(c). It
has a simpler pattern than for the polar case in
Fig.~\ref{fig:visAF}. The coherence of the initial matter-wave
exhibits collapse and revival modulations with a fast timescale of
$h/U_0$. Its frequency spectrum in panel (d) shows two dominant
frequencies, $(U_0-U_2)/h$ and $2(U_0-U_2)/h$, although other
frequencies with extremely small amplitudes do exist. For
$^{87}$Rb the dominant frequencies are $199$ and $398$ in units of
$U_2/h$.  As explained for $^{23}$Na, these frequencies appear
from the overlap of Fock states connected by the annihilation
operator $\hat{a}$~\cite{will10}. Here, the peak at $(U_0-U_2)/h$
appears due to the overlap of number states $|0,1,0 \rangle$ and
$|0,2,0 \rangle$. Similarly, the peak at $2(U_0-U_2)/h$ appears
due to number states $|0,2,0 \rangle$ and $|0,3,0 \rangle$.  As
with $^{23}$Na, finding these frequencies yields a method to
experimentally determine $U_2/U_0$. Nevertheless, since we need to
observe many oscillations, $\approx 200$ for $^{87}$Rb and
$\approx 80$ for $^{41}$K, this will be a challenging application
of quantum phase revival spectroscopy.

\subsection{Effects of magnetic field}

\begin{figure}
\vspace{-0.1cm}
\begin{center}
  \includegraphics[width=0.5\textwidth,angle=0]{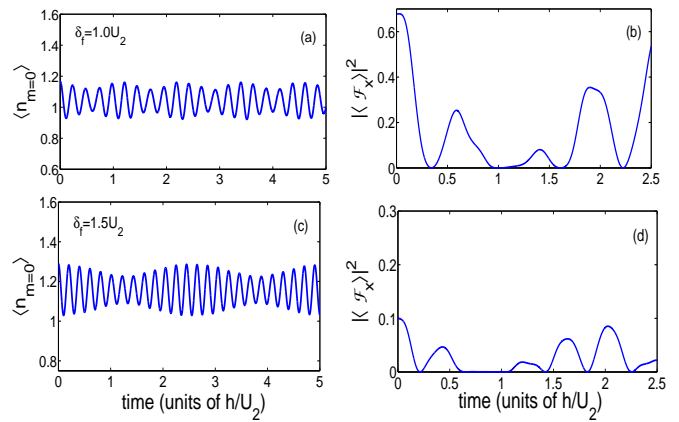}
\end{center}
\vspace{-0.6cm} \caption{\label{fig:Bferro} (color online) Effect
of quadratic Zeeman interactions for $^{87}$Rb dynamics for
$\delta=U_2$ and $1.5U_2$ for the top and bottom row respectively.
(a) and (c) shows spin mixing dynamics of the $m=0$ state, (b) and
(d) shows in-situ transverse magnetization $|\langle {\cal F}_x
\rangle|^2$ oscillations. Spin-mixing and $|\langle {\cal F}_x
\rangle|^2$ oscillations depend on a combination of $\delta$,
$U_0$, $U_2$ and the initial state. We see that the spin-mixing
amplitudes increase as we increase the field. On the other hand
transverse magnetization oscillations get weaker.}
\end{figure}

In this subsection we study the effects of the quadratic Zeeman
interaction on a ferromagnetic spinor. In Fig.~\ref{fig:Bferro} we
show the dynamics when the quadratic Zeeman shift is $\delta=U_2$
and $1.5U_2$ for the top and bottom row respectively. We plot spin
population $\langle {\hat n}_{m=0} \rangle$ and in-situ transverse
magnetization $|\langle {\cal F}_x\rangle|^2$ for $^{87}$Rb. The
parameters are $\langle {\hat n}\rangle =1.31$, $U_0/(zJ)=2$ and
$U_2=-0.005 U_0$ in the shallow lattice and $U_0=0.1\hbar\omega_f$
and $U_2=-0.005 U_0$ in the final deep lattice. In all the cases,
the oscillations are no longer periodic in $h/U_2$, but the values
of $\delta, U_0$ and $U_2$ combined with the initial state
composition influence the dynamics. We find that the spin-mixing
modulation amplitudes become large in the presence of a B-field
unlike the antiferromagnetic case in the previous section. The
oscillations also get faster, and the fast spin oscillations are
modified by an envelope of a slower modulation pattern which is
manifested in a beat-like pattern involving the two dominant
frequencies as clearly evident in Fig.~\ref{fig:Bferro}(c). As we
increase $\delta$, the population differences in the $m=0$ and
$m=\pm 1$ states become larger, and therefore a large spin-mixing
modulations can occur. Transverse magnetization $|\langle {\cal
F}_x \rangle|^2$ shown in (b) and (d) goes through partial
revivals and complete collapse. Faraday rotation spectroscopy can
be used to detect transverse magnetizations~\cite{lett09}. This
could give us a direct probe of the magnetic properties in our
quench set up, in addition to the coherence properties which we
can obtain through population and visibility revivals. We show in
panels (b) and (d) that the initial amplitude for $|\langle {\cal
F}_x \rangle|^2$ decreases for increasing B-field, but the number
of collapse sequences increases. When $\delta>2U_2$ the transverse
ferromagnetic superfluid turns into a longitudinal superfluid for
which $|\langle {\cal F}_x \rangle|^2$=0. In that regime
transverse magnetization is zero throughout the evolution, but
large amplitude spin-mixing modulations still occur.

\section{Signature of effective three-body interactions} \label{sec:three}

\subsection{Three-body interactions}

In our dynamical scheme, the evolution of the many-body state takes place
in a deep optical lattice with no tunnelling to the neighbors.  In such
a setting, even within the single-band Bose-Hubbard model, there are
effective three and higher-body interactions due to collision-induced
virtual excitations to higher bands or vibrational levels. For the spinless bosonic
case, such effective multi-body interactions have been predicted in
theory~\cite{tiesinga09}. A recent experiment~\cite{will10} monitored
matter wave collapse and revival dynamics for tens of oscillations and
observed the signature of higher-body effects in the visibility time
trace. A more accurate treatment of a multi-component system in a deep
lattice should therefore also incorporate higher-band induced multi-body
interaction terms.  Three-body interactions can be important for Efimov
physics~\cite{braaten07} and for many-particle systems giving rise to
novel and exotic phenomena~\cite{paredes07,pupillo09,mazza12}.

In a deep lattice, the minimum of the potential at a single site
can be approximated as a harmonic potential. For spin-1 bosons in
a single isotropic harmonic trap, the derivation of the effective
three-body interactions is given in the appendix. The effect can
be concisely represented by adding
\begin{equation}
H_{\rm 3B,eff} =
\frac{V_0}{6} \hat{n} \left(\hat{n}-1 \right) \left(\hat{n}-2 \right)
+ \frac{V_2}{6}  ({\vec{\cal F}}^2-2{\hat n})({\hat n}-2)
\label{onsite}
\end{equation}
to $H^{\rm final}$ in Eq.~(\ref{final}).  Here, as previously,
$\hat{n}=\hat{n}_{-1}+\hat{n}_0+\hat{n}_{1}$ is the on-site atom
number and $\vec{\cal F}=\sum_{m,m'}a^\dagger_{m} \vec{F}_{m m'}
a_{m'}$ is the on-site total angular momentum.  The Hamiltonian
term with strength $V_0$ is similar to the spin-0 effective
three-body interaction of Refs.~\cite{tiesinga09,will10} and only
depends on total particle number. The new Hamiltonian term with
strength $V_2$ depends intricately on the atomic spin. For a
harmonic potential with frequency $\omega_f$, $V_0$ is attractive
and equal to $V_0=-1.34 U_0^2/(\hbar \omega_f)$~\cite{tiesinga09}.
The strength $V_2$ satisfies $V_2=2(U_2/U_0)V_0$.  There is no
magnetic field dependence in the effective three-body terms.
Finally, we note that the perturbative effective three-body
Hamiltonian is only valid when the three-body interaction
strengths ($V_0, V_2$) are much smaller than the corresponding
two-body strengths ($U_0, U_2$).

The on-site effective Hamiltonian $H_{\rm 3B,eff}$ is diagonal in the
angular momentum basis states $|n,F,M \rangle$ with diagonal matrix
elements
\begin{equation}
  E_{\rm 3B,eff} =
\frac{V_0}{6} n (n-1 ) (n-2 ) + \frac{V_2}{6}  (F(F+1)-2 n)(n-2)\,.
    \label{spectrum3b}
\end{equation}
with the same restrictions as in Eq.~(\ref{energy}) on allowed
values of $F$. Equation~(\ref{spectrum3b}) is one of the key
results in this paper. This spectrum extends and generalizes the
spectrum of spin-mixing spinor hamiltonian as presented in
Ref.~\cite{law98} to the effective three-body case. As discussed
next, this helps us quantify and understand the effects of
three-body interactions on our dynamics scenario.

\subsection{Effects on Dynamics}

\begin{figure}
\begin{center}
    \includegraphics[width=0.45\textwidth,angle=0]{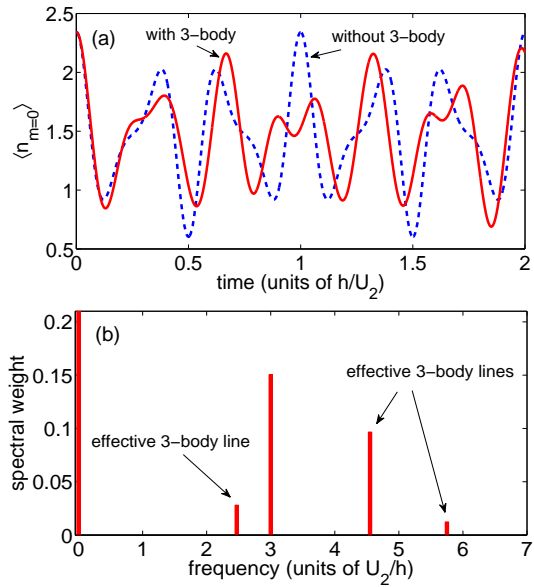}
\end{center}
\vspace{-1.7cm} \caption{\label{fig:af3b} Effects of effective
three-body interactions on the spin-mixing dynamics for a polar
($^{23}$Na) initial state. (a) The population $\langle {\hat
n}_{m=0}\rangle $ as a function of hold time for occupation
$\langle {\hat n} \rangle=2.35$ and $\delta=0$ with (solid line)
and without (dashed line) the effective three-body interaction,
(b) Frequency analysis of the time trace with effective three body
interactions reveals the presence of additional frequencies, which
could be used to determine the three-body interaction strength.}
\end{figure}

For spinless bosons in an optical lattice, a time of flight
measurement of the visibility dynamics determined the strength of
the effective multi-body interactions~\cite{will10,tiesinga11}.
Here for spin-1 bosons in an optical lattice, we show that the
effective three-body interaction effects can be observed directly
in the on-site population density -- in the spin oscillation
dynamics. This opens up the possibility that in situ measurements
such as quantum microscopes~\cite{greiner09,greiner11} and other
techniques~\cite{gemelke09} in lattices could be used to detect
effective multi-body interactions. Because of the more complex
nature of visibility patterns, we only analyze spin-mixing
population dynamics.

In Fig.~\ref{fig:af3b} we show the effects of the effective
three-body interactions in the spin-mixing dynamics of an initial
polar superfluid state of $^{23}$Na at $U_0/(zJ)=2$, $U_2 = 0.04
U_0$ and $\delta=0$. The occupation is $\langle {\hat n}
\rangle=2.35$, which highlights the three-body effects, and
$\langle {\hat m}\rangle=0$. The interaction strengths for the
deep lattice are $U_0 = 0.1 \hbar \omega_f$ and $U_2 = 0.04 U_0$,
so that $V_0 = -0.134 U_0$ and $V_2 = -0.268 U_2$. In an earlier
section we have seen that spin-mixing dynamics is controlled by
$U_2$. Here, we find that this spin-mixing scaling is also
influenced by the three-body interaction $V_2$. Figure
\ref{fig:af3b}(a) shows a comparison of the dynamics with and
without the three-body term. We see that the time traces start to
differ after the first oscillation, and the periodicity in $h/U_2$
is destroyed. A frequency analysis of the oscillations in panel
(b) elucidates the exact nature of the modulations. Without a
three-body term, strong features appear at $3 U_2/h, 5 U_2/h$, and
$7 U_2/h$. With a three-body term, additional frequencies appear
at $2.46 U_2/h, 4.55 U_2/h$, and $5.75 U_2/h$, which follow from
the energy differences of the three-body spectrum in
Eq.~(\ref{spectrum3b}).  Identification of any of the frequencies
gives us the $V_2$ coupling strength. For example, the peak at
$4.55 U_2/h$ arises as the initial state contains contributions of
angular momentum states $|3,1,0 \rangle$ and $|3,3,0 \rangle$
containing three atoms. These two states have an energy difference
of $5 U_2 +5 V_2/3$ whose signature is the $4.55 U_2/h$ frequency.

In experiments an unknown effective three-body strength $V_2$ can
be deduced by assigning several frequencies in the time trace. The
presence of all the other frequencies can be used to reduce error in
the measurement and verify the spectrum Eq.~(\ref{spectrum3b}). The
spectral weights and the values of the frequencies also give us
clue about the initial superfluid state. Control of initial
squeezing~\cite{will10,tiesinga11}, by varying $U_0/(zJ)$ in the shallow
lattice, can be used in such a way that some of the frequencies are more
dominant so to make detection easier.


\begin{figure}
\begin{center}
    \includegraphics[width=0.45\textwidth,angle=0]{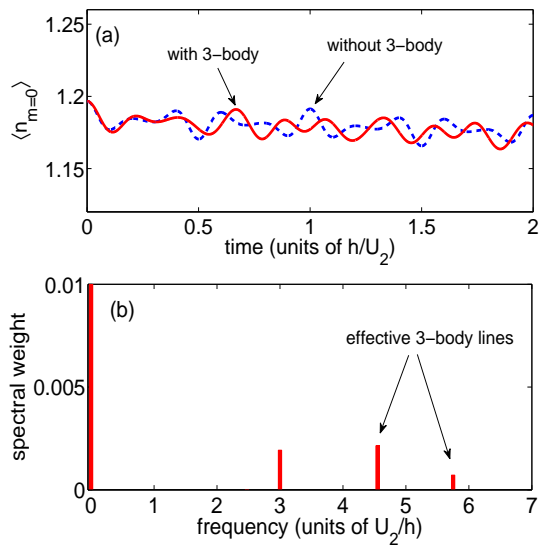}
\end{center}
\vspace{-2.3cm} \caption{\label{fig:ferro3b} Effects of effective
three-body interactions on the spin-mixing dynamics for a
ferromagnetic ($^{87}$Rb) initial state. (a) The population
$\langle {\hat n}_{m=0}\rangle$ as a function of hold time with
(solid line) and without (dashed line) the three-body interaction.
For $^{87}$Rb the oscillation amplitude is small. (b) Frequency
analysis reveals the presence of several frequencies, which are a
signature of induced three-body interactions; compared to
$^{23}$Na in Fig.~\ref{fig:af3b}, one frequency is missing for
$^{87}$Rb due to its ferromagnetic nature.}
\end{figure}

In Fig.~\ref{fig:ferro3b} we show the spin dynamics $\langle {\hat
n}_{m=0}\rangle$ for a ferromagnetic ($^{87}$Rb) initial state at
$U_0/(zJ)=2$, $\delta=0$, $\langle {\hat m}\rangle=0$, and $\langle
{\hat n}\rangle=2.35$ in the shallow lattice.  In the deep lattice we use
$U_0=0.1\hbar\omega_f$ and $U_2=-0.005 U_0$.  The spin-mixing amplitude
is not that prominent for $^{87}$Rb. Nevertheless, the influence of the
three-body interaction is discernible in our simulation.  The frequency
spectrum of the time trace shown in (b) makes it clear that new
frequencies emerge at $4.55 U_2/h$ and $5.75 U_2/h$ proving the existence
of three-body interactions and yielding a method to measure its strength.
The appearance of additional frequencies is similar to the $^{23}$Na case,
except that we do not observe a feature at $2.46 U_2/h$, which is due to a
coherence between four-atom angular momentum states $|4,0,0 \rangle$ and
$|4,2,0 \rangle$.  For the ferromagnetic initial superfluid the angular
momentum state $|4,0,0 \rangle$ is absent.  It is conceivable that initial
squeezing control of a ground state~\cite{will10,tiesinga11} or other
specific initial state preparations~\cite{law98,chapman05} can be used
to see larger amplitude spin-mixing oscillations to make the experimental
detection of three-body effects easier for ferromagnetic coupling.

\section{Conclusion} \label{sec:conclusion}

In this paper we performed a theoretical study of the dynamics of
spin-1 bosons in an optical lattice in a quench scenario where we
start from a ground state in a shallow lattice and suddenly raise
the lattice depth. We have shown that the ensuing spin-mixing and
visibility oscillations can be used as a probe of the initial
superfluid ground state. The spectral analysis of time evolution
reveals the Fock state composition of the initial state and
thereby its superfluid and magnetic properties. Analysis of
visibility oscillations, i.e. quantum phase revival spectroscopy,
further yields a method to determine the spin-dependent and
spin-independent interaction ratio $U_2/U_0$, which is an
important quantity for spinor gases. We treat both
antiferromagnetic (e.g. $^{23}$Na atoms) and ferromagnetic (e.g.
$^{87}$Rb) condensates. For ferromagnetic interactions the
spin-mixing oscillation amplitudes are small. When external
magnetic field cannot be ignored, the inclusion of a quadratic
Zeeman field is necessary, and we have quantified such dynamics.
We have shown that the presence of a magnetic field increases the
spin-mixing amplitude for a ferromagnetic condensate.

The Hamiltonian that more accurately describes the physics of the
final deep lattice comprises of two-body as well as effective
multi-body interactions, which arise due to virtual excitations to
higher bands. We derive the induced three-body interaction
parameters for spin-1 atoms in a deep harmonic well and show its
effect on the spin-mixing dynamics. We demonstrate that a
frequency analysis of the oscillations can detect the signature
and strength of the spin-dependent three-body interactions. We
stress our finding that the three-body interactions for spinor
atoms can be observed directly in the {\it in-situ} number
densities in addition to the time of flight visibility as observed
for spinless bosons~\cite{will10}.

Although there have been many theoretical studies for spin-1
bosons in an optical lattice, a many-body correlated ground state
has not yet been achieved experimentally. There are many
unexplored questions in that regard. Here, we have combined the
study of dynamics with optical lattice spinors to show how
non-equilibrium dynamics can be used as a probe for revealing
ground state properties and spinor interactions. There are other
dynamic scenarios that can give different perspectives on spinor
lattice physics, such as a quench from Mott insulator to
superfluid, evolution in a tunnel coupled lattice, to name a few.
Interplay of superfluidity, magnetism and strong correlations
makes this a rich system where study of quantum dynamics may lead
to a better understanding of collective phenomena.

\appendix

\section{Effective three-body spinor interactions}

In this appendix we derive the effective three-body spinor interaction
due to virtual excitations to excited trap levels of the isolated
sites of the deep optical lattice. The derivation closely follows
Refs.~\cite{tiesinga09,tiesinga11}. Atoms are held in the ground
state of a site, which we approximate by an isotropic 3D
harmonic-oscillator potential with frequency $\omega_f$.  Then the
Hamiltonian for spin-1 bosons is $H=H_0+V_0+V_2$ with
\begin{eqnarray}
  H_0 &=&  (\epsilon_{\alpha} +m^2\delta) a^\dagger_{m \alpha} a_{m \alpha} \\
   V_0 &=&
    \frac{1}{2} c_0  g_{\alpha\beta,\gamma\delta} a^\dagger_{k\alpha} a^\dagger_{l\beta} a_{k \alpha} a_{l\beta } \\
    V_2 &=& \frac{1}{2} c_2  g_{\alpha\beta,\gamma\delta}
 a^\dagger_{k\alpha } a^\dagger_{ l \beta} ({\vec F}_{km}\cdot
 {\vec F}_{ln}) a_{m \gamma} a_{n\delta }  \,,
\end{eqnarray}
where operators $a_{m \alpha}$ annihilate an atom in spin projection
$m=-1,0,1$ and 3D harmonic oscillator state $\alpha$.  We use the
convention that roman and greek subscripts represent spin projections and
harmonic oscillator states, respectively. Repeated indices are summed
over. The single-particle energies $\epsilon_\alpha$ and $m^2\delta$
are the harmonic oscillator and quadratic Zeeman energies, respectively.
For the ground state $\alpha=0$. We choose $\epsilon_0=0$ and
will require that $\epsilon_\alpha-\epsilon_0\ll \delta$ for $\alpha\neq 0$.
The spin-independent and spin-dependent atom-atom interactions $V_0$
and $V_2$ have ``bare-coupling'' strengths $c_0$ and $c_2$,
respectively.  The vector ${\vec F}=(F_x,F_y,F_z)$ are spin-1 matrices. Each
symmetric real coefficient $g_{\alpha\beta,\gamma\delta}$ is a 3D integral
over the product of the four oscillator wavefunctions $\alpha$,
$\beta$, $\gamma$, and $\delta$.  In this appendix the two atom-atom
interactions are explicitly normal-ordered in order to facilitate the
derivation.

We can now derive the effective Hamiltonian with atoms in the
lowest oscillator state using degenerate perturbation theory with
zeroth-order Hamiltonian $H_0$ and perturbation $V_0+V_2$.  First,
we define for every spatial mode $\alpha$ the spin wavefunction
$|\{n_{-1},n_0,n_1\}_\alpha\rangle$ with $n_m=0,1,2,\dots$ atoms in spin
state  $m$.  The ground states $P$ are formed by the orthonormal
basis functions $| g \rangle = | \{n_{-1},n_0,n_1\}_0,  0_{\alpha\ne
0}\rangle$ for any value of $n_m$ and where $0_{\alpha\ne 0}$ indicates
that there are no atoms in excited spatial modes.  Their energy is
$E_g=(n_{-1}+n_{1})\delta$. Excited states $|e\rangle$ with energy $E_e$
are states where at least one atom occupies a $\alpha\ne 0$ spatial mode.

We reproduce the Hamiltonian in Eq.~(\ref{final}) in first-order
perturbation theory once we make the assignment $U_0=c_0 g_{00,00}$ and
$U_2=c_2 g_{00,00}$. To second-order in degenerate perturbation theory
the matrix element for ground states $|g'\rangle $ and $|g\rangle$ is
\begin{eqnarray}
  \langle g' |\delta H^{(2)} | g\rangle &=&
      \frac{1}{2}\sum_{e\ne P}
         \langle g'| V_0+V_2 | e\rangle
      \left\{  \frac{1}{E_g-E_e}  \right.
\\
 &&  \quad\quad\quad
        \left. +
           \frac{1}{E_{g'}-E_e} \right\}\langle e | V_0+V_2 | g\rangle
    \,. \nonumber
\end{eqnarray}

The sum over excited states can be evaluated by realizing that
only states $|e\rangle\propto a^\dagger_{k\alpha }
a^\dagger_{l\beta } a_{m 0} a_{n 0} |g\rangle$ with
$\alpha\beta\ne 00$ and $k+l=m+n$ contribute as both $V_0$ and
$V_2$ conserve atom number and magnetization, and can only change
the state of two atoms at the same time, i.e. states $|e\rangle$
are those with at most two atoms in the higher trap levels.  By
inspection, we then realize that the energy differences
$E_{e}-E_g\approx\epsilon_\alpha+\epsilon_\beta$ are independent
of the total number of atoms in the ground states and to good
approximation are also independent of the quadratic Zeeman energy,
as $\delta\ll\epsilon_\alpha$ for $\alpha\ne 0$.   Similar
expressions hold for $E_e-E_{g'}$.

Inserting the (normalized) expression for $|e\rangle$ and performing the sums over
$|e\rangle$ as well as those appearing in the potentials $V_0$ and $V_2$ we first find
\begin{eqnarray}
   \lefteqn{ \langle g' |  \delta H^{(2)}  | g \rangle =  -
\sum_{\alpha\beta\ne 00,\mu\nu} g_{00,\mu\nu} \frac{1}{\epsilon_{\alpha}+\epsilon_\beta}
g_{\alpha\beta,00} }\nonumber\\
    && \times \left\{
   \frac{1}{4} c_0^2
                   \langle g'|
              a^\dagger_{m 0} a^\dagger_{n 0} a_{m\mu } a_{n\nu  }
              a^\dagger_{k\alpha } a^\dagger_{l\beta } a_{k 0} a_{l 0}
| g \rangle \right. \label{first} \\
    && \left. +\frac{1}{2} c_0 c_2
         \langle g'|
a^\dagger_{p 0} a^\dagger_{o 0} ({\vec
F}_{pm}\cdot
 {\vec F}_{on}) a_{m\mu} a_{n\nu}
   a^\dagger_{k\alpha} a^\dagger_{l\beta} a_{k 0} a_{l 0} | g \rangle \right\} \,,
       \nonumber
\end{eqnarray}
where the remaining sums over trap levels have been made explicit,
repeated roman indices are still summed over, and we have omitted the
contribution proportional to $V_2$ times $V_2$ as the spin-dependent
interaction strength is an order of magnitude smaller than the
spin-independent one.

By normal ordering  the creation and annihilation operators in
Eq.~(\ref{first}) and using that ground states $P$ contain no atoms in
excited trap levels  we find
\begin{eqnarray}
  \langle g'| \delta H^{(2)} | g\rangle &=&\langle g' | \delta H_{\rm 2B} + H_{\rm 3B} | g\rangle \,,
\end{eqnarray}
where
\begin{eqnarray}
 \delta H_{\rm 2B}  &=& Z_2 \Bigl\{
       - \frac{1}{2} c_0^2
        a^\dagger_{k 0} a^\dagger_{l 0} a_{k 0} a_{l 0}
     \\
  && \quad\quad\quad\quad  - c_0 c_2
             a^\dagger_{m 0} a^\dagger_{n 0} ({\vec F}_{mk}\cdot {\vec F}_{nl}) a_{k 0} a_{l 0}
               \Bigr\}  \nonumber
\end{eqnarray}
is a correction to the pair-wise two-body interaction and
\begin{eqnarray}
 H_{\rm 3B}  &=& 
           Z_3
    \left\{
      -c_0^2
        a^\dagger_{k 0} a^\dagger_{l 0} a^\dagger_{m 0} a_{k 0} a_{l 0} a_{m 0}
            \right.  \\
   && \quad\quad\quad  \left. -2 c_0 c_2
       a^\dagger_{o 0} a^\dagger_{m 0} a^\dagger_{n 0} ({\vec F}_{m k}\cdot
        {\vec F}_{n l})  a_{k 0} a_{l 0} a_{o 0}
             \right\}     \nonumber
\end{eqnarray}
is an effective three-body interaction.
Here
\begin{equation}
  Z_2= \sum_{\mu\nu\ne 00} g_{00,\mu\nu} \frac{1}{\epsilon_{\mu}+\epsilon_\nu} g_{\mu\nu,00}
\end{equation}
and
\begin{equation}
  Z_3 = \sum_{\mu\ne 0} g_{00,\mu0} \frac{1}{\epsilon_{\mu}} g_{\mu0,00} \,.
\end{equation}

The sums in $Z_2$ diverge and must be regularized and renormalized
\cite{tiesinga11,fetter}.  That is we require that the bare
coupling constant $c_0$ is defined such that, by combining the
first- and second-order contributions, $c_0g_{00,00}- c_0^2 Z_2$
is finite and equal to $U_0$. Similarly, we require that
$c_2g_{00,00}-2c_0c_2Z_2$ is finite and equal to $U_2$.

The sums in coefficient $Z_3$ of the effective three-body
Hamiltonian do converge.
Hence, we redefine the effective three-body interaction as
\begin{eqnarray}
          H_{\rm 3B} &=& \frac{1}{6} V_0 \, a^\dagger_{k 0} a^\dagger_{l 0}
 a^\dagger_{m 0} a_{k 0} a_{l 0} a_{m 0} \\
     && \quad + \frac{1}{6} V_2 \,
             a^\dagger_{o 0} a^\dagger_{m 0} a^\dagger_{n 0}
                  ({\vec F}_{mk}\cdot {\vec F}_{nl})
         a_{k 0} a_{l 0} a_{o 0} \,, \nonumber
\end{eqnarray}
where, consistent within our second-order perturbative calculation,
\[
     V_0=-6 U_0^2 \, \frac{1}{g_{00,00}^2} Z_3
\ \ {\rm and}\ \
     V_2=-12 U_0 U_2 \, Z_3 = 2 \frac{U_2}{U_0} V_0
\]
are the spin-independent and spin-dependent three-body interaction
strength, respectively.  For a spherically symmetric harmonic oscillator
$V_0=-1.34\dots U_0^2/(\hbar\omega_f)$~\cite{tiesinga09}.  Since $U_2>0$
for Na atoms, $V_2<0$, while for $^{87}$Rb $U_2<0$ and thus $V_2>0$.

The two- and three-body interaction can be rewritten in terms of the angular
momentum operators $\vec {\cal F}=(\hat {\cal F}_x, \hat {\cal F}_y, \hat
{\cal F}_z)$ defined in Eq.~(\ref{hamil}) following Ref.~\cite{law98}.
By combining the quadratic Zeeman interaction as well as the two-body and effective three-body interaction we find our final result
\begin{eqnarray*}
      H_{\rm eff} &=& \delta \sum_m m^2 a^\dagger_m a_m+\frac{1}{2}U_0 {\hat n}({\hat n}-1)+
               \frac{1}{2}U_2 (\vec {\cal F}^2-2{\hat n}) \\
   && + \frac{1}{6} V_0 {\hat n}({\hat n}-1)({\hat n}-2)+
                   \frac{1}{6}   V_2 (\vec {\cal F}^2-2{\hat n}) ({\hat n}-2)  \,,
\end{eqnarray*}
where we suppressed the ground-state oscillator index, and $\vec
{\cal F}^2$ and the number operator $\hat n$ commute.

\begin{acknowledgments}
KWM would like to thank Richard Scalettar and Hulikal
Krishnamurthy for discussions during early stages of this work. We
acknowledge support from the US Army Research Office under
Contract No. 60661PH and the National Science Foundation Physics
Frontier Center located at the Joint Quantum Institute.
\end{acknowledgments}

{}

\end{document}